\documentclass[final,5p,times,twocolumn]{elsarticle}

\usepackage[utf8]{inputenc}

 \usepackage{graphics}

\usepackage{amssymb}
\usepackage{hyperref}
\hypersetup{colorlinks=true,%
           linkcolor=black,%
           anchorcolor=black,%
           citecolor=black,%
           filecolor=black,%
           menucolor=black,%
           urlcolor=black}

\usepackage{lineno}

\usepackage{subfigure}

\usepackage{stfloats}

\newcommand{\neqcm}{\ensuremath{\mathrm{n}_{\mathrm{eq}}/\mathrm{cm}^2}}
\newcommand{\mum}{\ensuremath{\mu}m}

\newcommand{\Sr}{\ensuremath{^{90}}Sr}

\journal{Nuclear Instruments and Methods A }

\begin{document}

\begin{frontmatter}




\title{Recent Results of the ATLAS Upgrade Planar Pixel Sensors R\&D Project}
\author{Philipp~Weigell\corref{cor1}\fnref{label1}}
\ead{Philipp.Weigell@mpp.mpg.de}
\cortext[cor1]{Corresponding author}
\author{on behalf of the ATLAS Planar Pixel Sensors R\&D Project}
\address[label1]{Max-Planck-Institut f\"ur Physik, F\"ohringer Ring 6, D-80805 M\"unchen, Germany}

\begin{abstract}
To extend the physics reach of the LHC experiments, several upgrades to the
accelerator complex are planned, culminating in the HL-LHC, which eventually leads to
an increase of the peak luminosity by a factor of five to ten compared to the
LHC design value.

To cope with the higher occupancy and radiation damage also the LHC experiments
will be upgraded. The ATLAS Planar Pixel Sensor R\&D Project is an
international collaboration of 17 institutions and more than 80 scientists,
exploring the feasibility of employing planar pixel sensors for this scenario.

Depending on the radius, different pixel concepts are investigated using
laboratory and beam test measurements. At small radii the extreme radiation 
environment and strong space constraints are addressed with very thin pixel 
sensors active thickness in the range of (75--150)\,\mum{}, and the development of 
slim as well as active edges. At larger radii the main challenge is the 
cost reduction to allow for instrumenting the large area of (7--10)\,m$^2$. To 
reach this goal the pixel productions are being transferred to 6 inch production 
lines and more cost-efficient and industrialised interconnection techniques are 
investigated. Additionally, the n-in-p technology is employed, which requires less 
production steps since it relies on a single-sided process.

An overview of the recent accomplishments obtained within the ATLAS Planar Pixel Sensor R\&D Project
is given. The performance in terms of charge collection and tracking
efficiency, obtained with radioactive sources in the laboratory and at beam
tests, is presented for devices built from sensors of different vendors
connected to either the present ATLAS read-out chip FE-I3 or the new Insertable B-Layer
read-out chip FE-I4. The devices, with a thickness varying between 75\,\mum{} and 300\,\mum, 
were irradiated to several fluences up to $2\cdot10^{16}$\,\neqcm. Finally, 
the different approaches followed inside the collaboration to achieve slim 
or active edges for planar pixel sensors are presented.
\end{abstract}


\end{frontmatter}



\section{Upgrades Roadmap}
\label{sec:introduction}
Presently, the ATLAS pixel detector \cite{pixelelectronics} comprises three barrel layers located at radii between 50.5\,mm and 122.5\,mm as well as three end-cap discs on each side of the detector. In total about 80 million read-out channels are distributed on 1744 pixel modules. Each module is composed of a 250\,\mum{} thick n-in-n planar silicon sensor interconnected via the solder bump bonding technique \cite{Fritzsch2011189} to 16 FE-I3 read-out chips \cite{Peric2006178}, featuring pixel pitches of 50\,\mum{}\,$\times$\,400\,\mum{}. Sensors and read-out chips are specified up to a fluence of $10^{15}$\,\neqcm{} (1\,MeV neutrons) or a dose of 500\,kGy.

\begin{figure}[ht]
\centering
\includegraphics[width=\columnwidth]{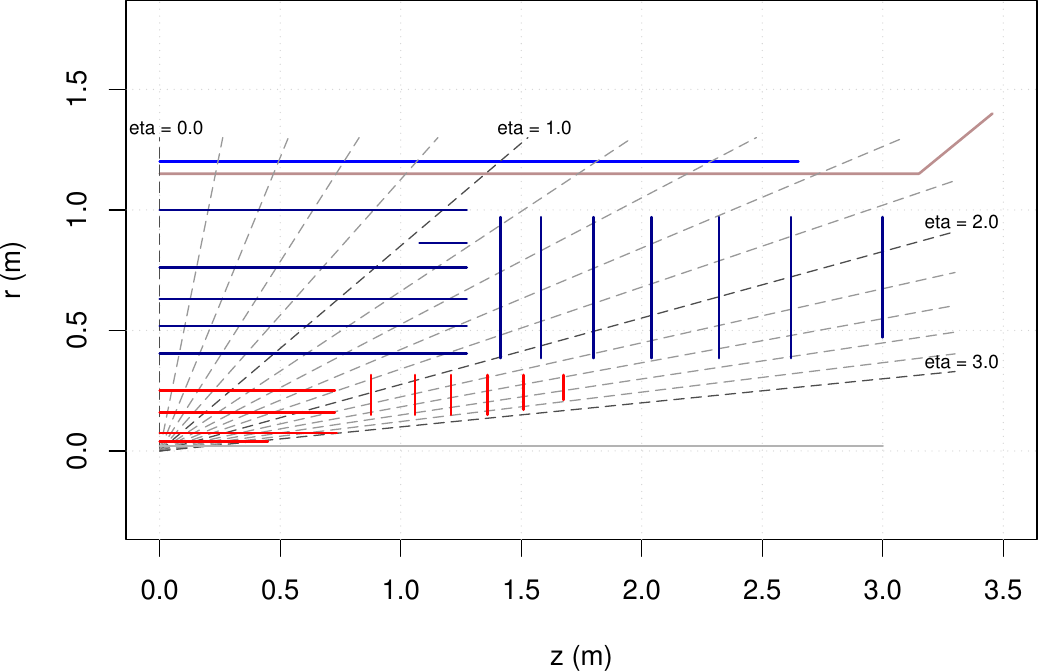}
\caption{Baseline layout of the new inner detector for the Phase II upgrade \cite{LOI_II}. Pixel (strip) layers and discs are indicated in red (blue).}
\label{fig:Baseline}
\end{figure}

To increase the physics reach of the LHC programme, it is foreseen to upgrade the accelerator chain in three dedicated long shutdowns (LS), followed by longer data-taking phases, called phase 0, I, and II. While increasing the beam energy to its design value, the peak luminosity will increase eventually up to (5--8)$\cdot10^{34}$\,cm$^{-2}$s$^{-1}$ \cite{Lumi}. Each LS will be mirrored by upgrades to the ATLAS detector to cope with the increased luminosity. This paper will focus on the upgrades of the pixel detector, only. The first LS starts beginning of 2013 and lasts until the end of 2014; it will lead to an approximately fourfold increase in luminosity. In the ATLAS detector a new fourth pixel layer will be mounted on a new smaller beam pipe at a radius of 32\,mm. This is referred to as the Insertable B-Layer (IBL) \cite{IBL-TDR}. The smaller radius inhibits overlapping modules in $z$ as employed in the present ATLAS pixel detector. Thus, the active fraction had to be increased, using a new design of the n-in-n sensors discussed in Section\,\ref{sec:slimedge}. Given the harsher radiation environment and the higher occupancy a new read-out chip, the FE-I4 \cite{GarciaSciveres2010}, was developed, which is specified up to a received fluence of $5\cdot10^{15}$\,\neqcm. The pixel cell size was reduced to 50\,\mum{}$\,\times\,$250\,\mum{} and the number of pixel cells increased from 2880 to 26880. While the upgraded pixel detector is believed to retain sufficient tracking capabilities after the second LS, which starts around 2017, during the third LS from 2021 to 2022 a major upgrade of the entire inner tracking system is planned. The replacement of the tracking detector is required given the foreseen fluences in phase II of up to $2\cdot10^{16}$\,\neqcm{} in the innermost layer, along with the very high occupancies, calling for higher granularity and a new generation of read-out chips for the inner layers. The current baseline layout planned is depicted in Figure\,\ref{fig:Baseline}.
The barrel consists of four pixel layers, with a minimal radius around 39\,mm and a maximal radius around 250\,mm. Six pixel discs are foreseen for the forward region, i.\,e.\ at a pseudorapidity of about $1.8\leq|\eta|\leq2.8$. Depending on performance simulations, it is planned to increase the radius even further, or to add an additional fifth pixel layer. 

\section{The ATLAS Planar Pixel Sensor R\&D Project}
The scope of the ATLAS Planar Pixel Sensor R\&D Project is to evaluate and improve the performance of planar pixel sensors for these detector upgrades as well as to determine their operation conditions in this high luminosity environment. Planar pixel sensors are used widely already in present high energy physics experiments and have very well established manufacturing processes with very high yield and low costs. Still, at the irradiation levels expected in the inner pixel layers at the HL-LHC, they require high bias voltages and pose stronger constraints on the cooling systems, especially when comparing to emerging sensor technologies like 3D sensors \cite{Parker3D}, which still are affected by lower yield and higher production costs, but made significant progress over the last few years and will be used in the ATLAS IBL upgrade \cite{IBL_Proto,Grenier201133,DaVia2012321}. Besides radiation hardness studies, geometry optimization and cost reduction are the key topics, which will also be covered in this paper. To achieve these goals, sensor productions from CiS \cite{CiS}, FBK \cite{FBK}, HPK \cite{HPK}, Micron \cite{Micron}, MPI-HLL \cite{MPI-HLL}, and VTT \cite{VTT} are used. To investigate the properties for realistic scenarios, irradiations with various particle types and energies are used, i.\,e.\ reactor neutrons (JSI) \cite{Snoj}, and protons with 26\,MeV (KIT) \cite{KIT}, 800\,MeV (LANSCE) \cite{LANSCE}, and 24\,GeV (CERN PS) \cite{CERNPS}. 
The samples are then measured in the laboratory using radioactive sources and in beam test at the CERN SPS and DESY, where the EUDET beam telescope \cite{eudet} is employed. The experimental measurements are supported by TCAD simulations. 

\section{Phase II Requirements}
Investigations within the ATLAS Planar Pixel Sensor R\&D Project are focusing towards the phase II upgrade of the ATLAS inner tracking system. The results presented in the following are grouped according to the radius they are most relevant for.

\subsection{Phase II -- Outer Layer}
The outer pixel layers drive the total area to unprecedented values of about (7--10)\,m$^{2}$, thus cost effective modules are mandatory. To achieve this, cost-reduction for the sensors as well as for the interconnection is envisaged. Since the latter one is mostly driven by the number of tiles to be interconnected, it is foreseen to use large area sensors, which are then interconnected to four or even six FE-I4 chips. Three productions are ongoing at the moment which include either four chip sensors or have pairs of two-chip sensors placed close on the wafer, such that they can be diced as a four-chip sensor. The sensors were designed by the KEK group \cite{KEKquad}, the University of Liverpool group, and the MPP/HLL group, and are produced by HPK, Micron, and CiS respectively. 

\subsubsection{Performance of n-in-p Pixel Detectors}
Since the pn-junction is on the pixel implantation side in n-in-p sensors, the guard rings can be placed on the front-side as well, and thus patterned processing is only needed on a single side. Consequently, no masks and no alignment for back-side processing are needed, lowering the cost and enabling the use of more foundries for processing. Furthermore, the lack of patterned back-side implantations eases subsequent handling and testing. A possible problem connected to the n-in-p geometry is related to the high voltage present at the edges at the front-side, transferred from the backside through crystal damages along the sensor sides. Since the sensor edge region is facing the read-out chip, which is at ground potential, at a distance of O(10\,\mum{}), destructive electric discharges are possible and were observed \cite{RoheSparks}. To prevent this, three different methods were investigated.  In the first approach a 3\,\mum{} Benzocyclobutene (BCB) \cite{BCB1,BCB2} passivation layer on the sensor surface was used and no destructive discharge observed up to a bias voltage of 1\,kV \cite{NinPpaper}. As alternatives two post processing approaches, employing silicon adhesive and Parylene-C have been investigated. Up to 1\,kV no destructive discharges were found when using a full silicon adhesive encapsulation of the module. Modules encapsulated with Parylene-C have been tested up to 650\,V, and again no destructive discharges were seen. In none of the three approaches a degradation of the high voltage insulation was observed after irradiation. 

An extensive radiation programme, using sensors from three different productions, was conducted to investigate the performance of n-in-p pixel modules after high received fluences up to $10^{16}\,\neqcm$. The first production using designs by the MPP/HLL group was processed at CiS on 285\,\mum{} thick wafers \cite{NinPpaper}. The other two productions yielded 150\,\mum{} thick sensors and were conducted by the KEK group in collaboration with HPK \cite{RyoHiro,UnnoHiro} and by the MPP/HLL group \cite{AnnaJapan,WeigellPhD}.

\begin{figure}[ht]
\centering
\includegraphics[width=0.98\columnwidth]{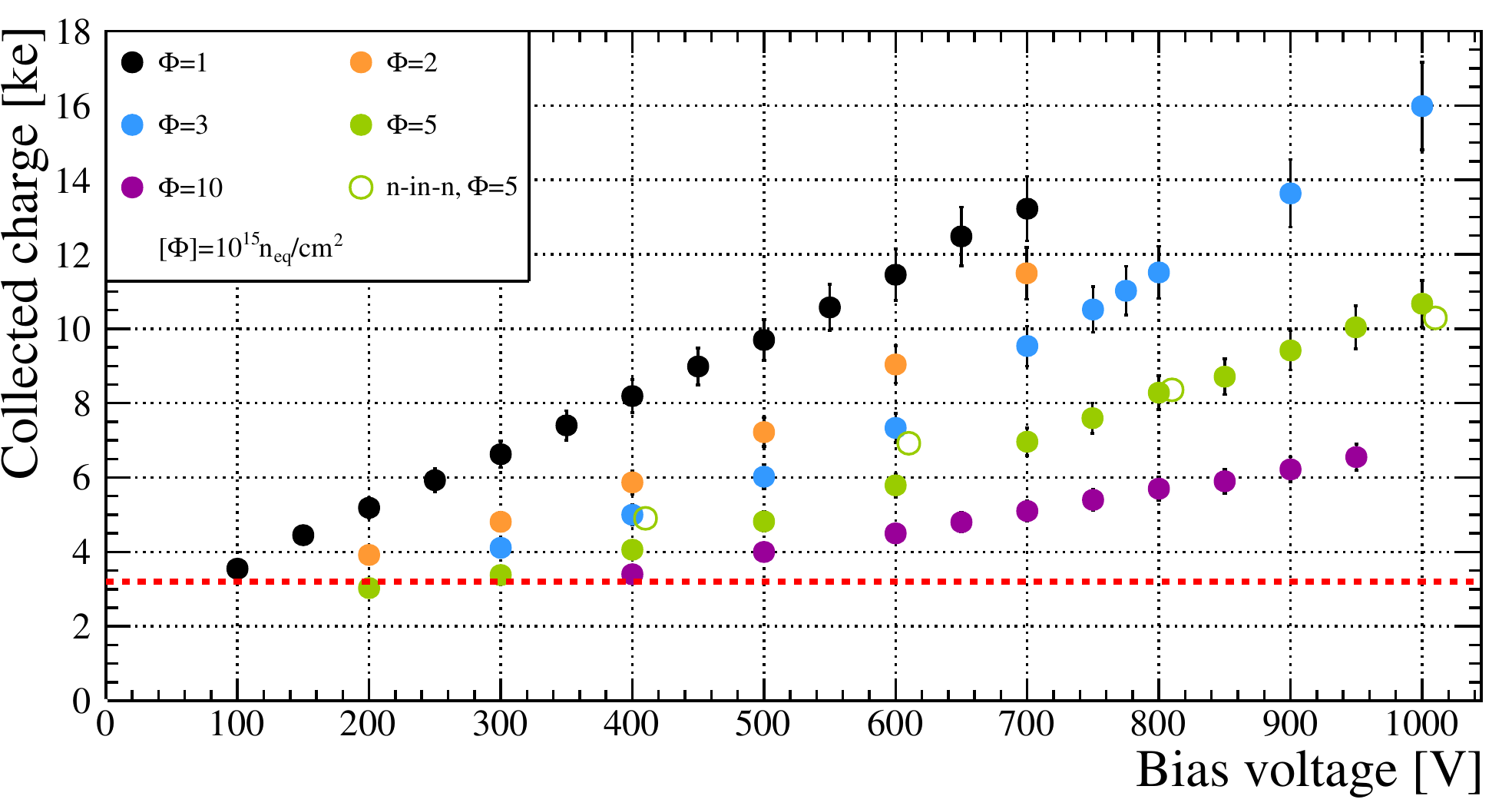}
\caption{MPV of collected charge for neutron irradiated n-in-p modules, obtained from $^{90}$Sr source measurements, as a function of the bias voltage. The uncertainties are fully correlated and account for charge calibration uncertainties. For a discussion please refer to \cite{Ale,WeigellPhD}. The results from the n-in-n module are from \cite{Altenheiner11}; for better visibility, these points are drawn horizontally displaced. The dotted red line indicates the threshold at 3.2\,ke.}
\label{fig:CisNinPnirrad}
\end{figure}

In Figure\,\ref{fig:CisNinPnirrad} the most probable values (MPVs) of the collected charges are summarised as a function of the applied bias voltage for various received fluences for the modules from the CiS production. The MPV of the collected charge rises with bias voltage and decreases with received fluence. For all modules the collected charge exceeds the threshold of 3.2\,ke by a factor of 2 with bias voltages below 1\,kV, indicating a good hit efficiency. For comparison measurements using an n-in-n irradiated module \cite{Altenheiner11} are shown as well. 

With beam test measurements at the CERN SPS employing 120\,GeV pions and at DESY using 4\,GeV positrons the hit efficiency was determined as a function of the bias voltage and received fluence for several modules from the different productions.
\begin{figure*}[b]
\centering
\subfigure[]{
\includegraphics[width=0.98\textwidth]{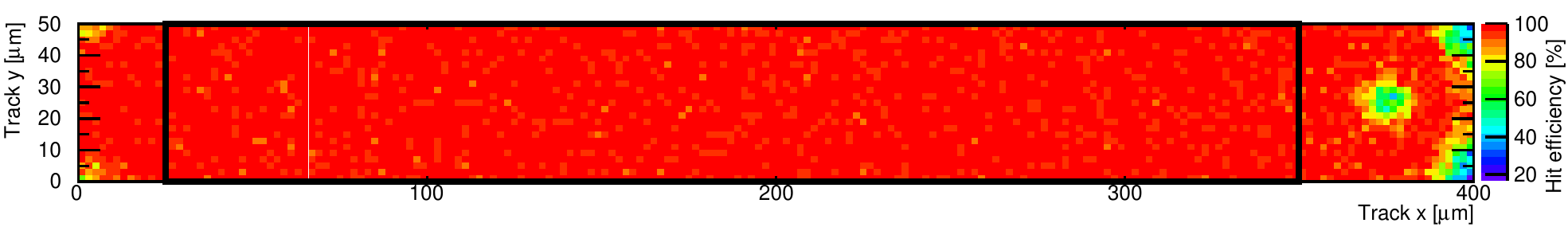}
\label{fig:pixeff1e16}
}
\subfigure[]{
\includegraphics[width=0.882\textwidth,angle=180]{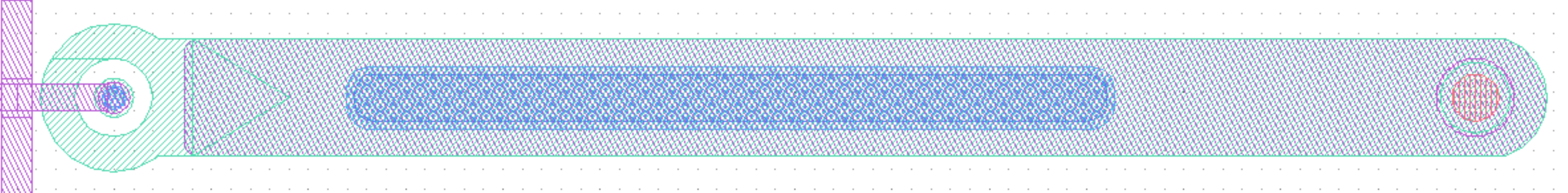}
\label{fig:pixelgeometry}
}
\caption{\subref{fig:pixeff1e16} Mean hit efficiency as a function of the impact point predicted by the beam telescope for a FE-I3 module with a thickness of 285\,\mum{} irradiated to a fluence of $10^{16}$\,\neqcm{} and operated at a bias voltage of 600\,V for a threshold of 2\,ke. \subref{fig:pixelgeometry} Design of a single pixel. The implantation extends over the entire structure shown, and has a ring shaped opening at the punch through bias dot displayed on the right side. The metal layer, covering most of the implant, is shown as a large rectangle with rounded corners on the left side. The T-shaped structure at the far right end comprises the metal lines, connecting the bias dot to the bias ring. The opening in the nitride and oxide layers is displayed as the rectangle in the centre of the pixel. The small circle at the left end of the pixel is the opening in the passivation, where the pixel will be connected with bump-bonding.}
\label{fig:Lub1e16}
\end{figure*}
For a module from the CiS production, irradiated to a fluence of $10^{16}$\,\neqcm{}, and operated at a moderate bias voltage of 600\,V for a threshold tuned to 2\,ke the mean hit efficiency was determined to be still as high as $(97.2\pm0.3)$\,\%. In Figure\,\ref{fig:Lub1e16} the hit efficiency is depicted as a function of the impact point predicted by the beam telescope projected into one pixel cell together with the design of the pixel cell. The main hit efficiency losses occur in the regions of the bias dot and in the corners. The former is because the implant in the bias dot is not connected to the read-out, the latter due to charge sharing among several pixels, bringing all participating pixel cells below threshold. Anyhow, both effects are only relevant for perpendicular impinging particles, occurring only for very few parts of a high energy physics experiment. Therefore, the quoted hit efficiency has to be understood as a lower bound. If only the central part is considered, indicated by the box in Figure\,\ref{fig:pixeff1e16}, the hit efficiency is $(98.1\pm0.3)$\,\%. Further details can be found in \cite{Ale,WeigellPhD}.

\begin{figure}[hbt]
\centering
\subfigure[]{
\includegraphics[width=0.98\columnwidth]{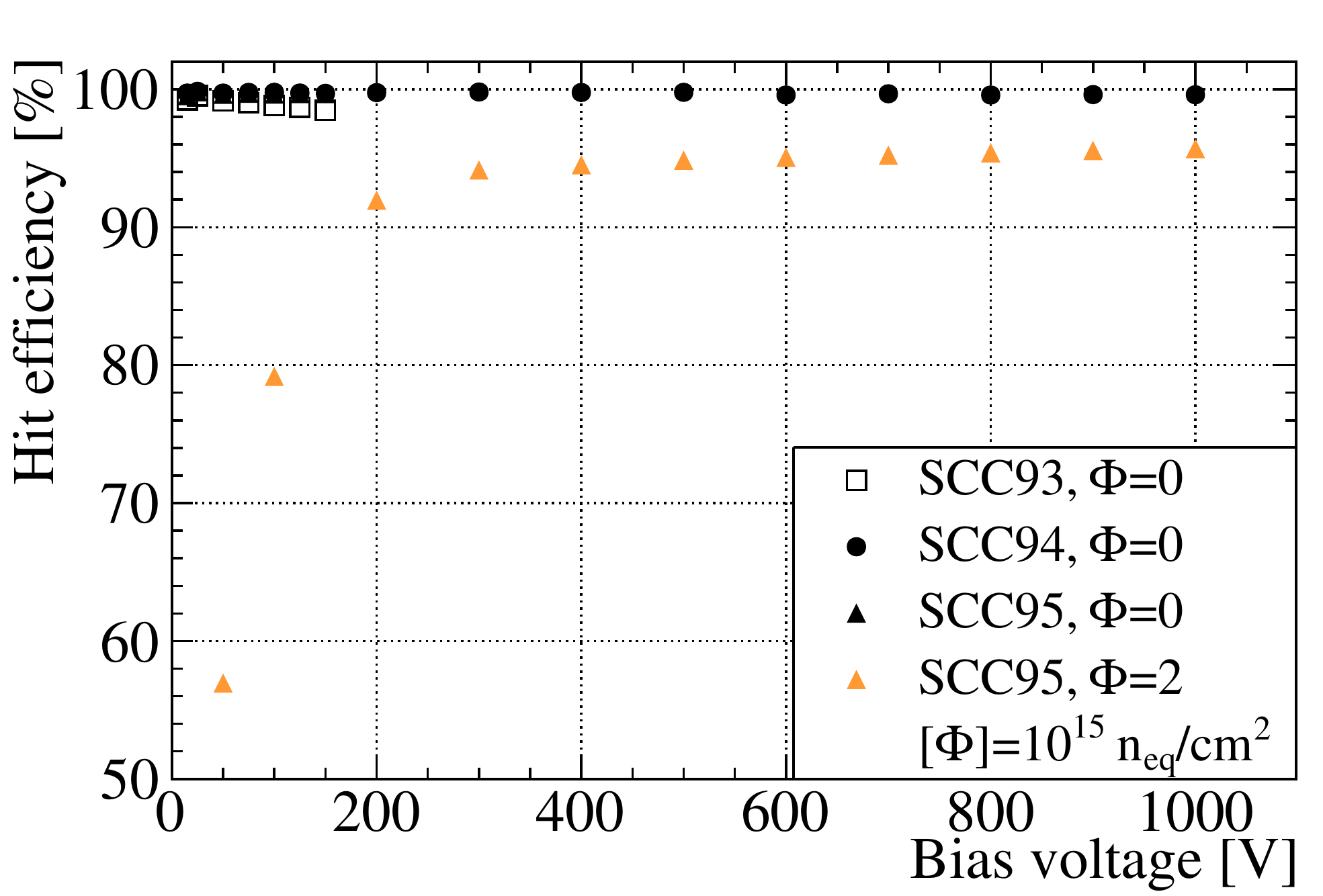}
\label{fig:KeKeff}
}
\subfigure[]{
\includegraphics[width=0.98\columnwidth]{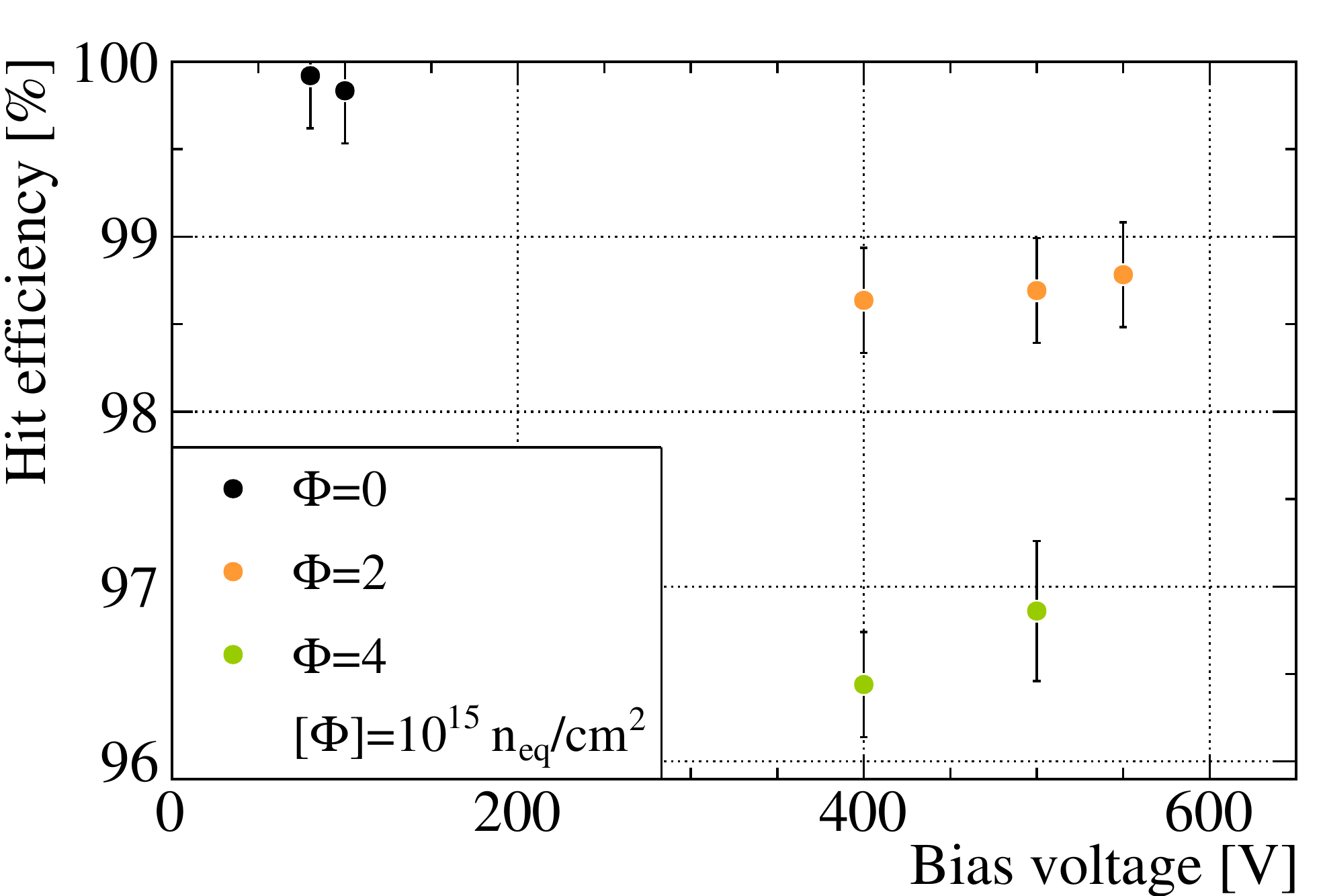}
\label{fig:MPPeff}
}
\caption{Hit efficiency as a function the applied bias voltage for n-in-p sensors with a thickness of 150\,\mum{} connected to the FE-I4 read-out chip before and after irradiation. In \subref{fig:KeKeff} the results for the HPK modules are shown before and after irradiation to a fluence of $2\cdot10^{15}$\,\neqcm with protons at KIT. \subref{fig:MPPeff} summarises the results for the MPI-HLL modules. The received fluence is indicated by the colour. The given uncertainties are systematic.}
\label{fig:FEI4eff}
\end{figure}

In Figure\,\ref{fig:FEI4eff} the hit efficiencies for the modules based on the 150\,\mum{} thick sensors and FE-I4 read-out chips are summarised as a function of the bias voltage and the received fluence. Before irradiation (black points) all modules show an excellent performance with a hit efficiency above 99.7\,\%. After irradiation to a fluence of $2\cdot10^{15}$\,\neqcm{} with low energetic protons at KIT the HPK modules exhibit a stable performance up to 1\,kV. The MPI-HLL modules were irradiated at KIT to $2\cdot10^{15}$\,\neqcm{} and at LANSCE to $4\cdot10^{15}$\,\neqcm. Also here a good hit efficiency is found already at low bias voltages, with the main losses again occurring in the the above mentioned regions. For further details please refer to \cite{RyoHiro,AnnaJapan}.

\subsection{Phase II -- Inner Layer}
For the inner layers of the pixel detector foreseen in the phase II upgrade the requirements are different. The maximum expected charged particle density of $1.07\cdot10^{-1}$\,cm$^{-1}$ per proton-proton collision \cite{dawson} makes the best achievable resolution mandatory and thus a reduction of the $R\phi$-pitch to 25\,\mum{} and the $z$-pitch to 150\,\mum{} for the sensor as well as for the read-out chip is envisaged. By this increase in granularity the charged particle density is reduced by a factor of three to $4\cdot10^{-6}$ per pixel and proton-proton collision when compared to the IBL pixel geometry. While implementing such small pitches can be achieved for the read-out chip by switching to 65\,nm technology \cite{MauriceJapan}, they are not yet achievable on a large scale within the interconnection process. Thus, improvements to the current technologies are needed, or new techniques like Solid Liquid Inter-Diffusion (SLID) \cite{AnnaJapan,PhilippAbery}, offering lower pitches down to approximately 20\,\mum{}, have to be used. To investigate the performance of 25\,\mum{} wide pitches already today, special designs are currently produced. One by the KEK group on the HPK production lines, and one by the groups of the University of Liverpool and the University of Glasgow at Micron. In these designs, the pixel pitch is reduced to 25\,\mum{} in $R\varphi$, by merging pixel implants of two columns in the $z$ direction, creating a pitch of 500\,\mum{}. The bump pads are arranged with a metal routing layer compliant with the FE-I4 pixel cell geometry of $50\times250$\,\mum$^2$. A complementary approach to achieve the best possible resolution is to decrease the mounting radius of the pixel layer. This implies that no overlap of the modules in $z$ is possible for geometric reasons, as in the IBL. To ensure a full coverage, within the ATLAS Planar Pixel Sensor R\&D Project several methods are investigated to achieve slim edge regions. These will be discussed in Section\,\ref{sec:slimedge}.  Being so close to the 
interaction point radiation hardness up to $2\cdot10^{16}$\,\neqcm{} will be needed and a reduction of multiple scattering is mandatory. While the radiation hardness of the n-in-p modules was discussed was already discussed in the preceding section, it will be presented for the n-in-n modules in Section\,\ref{sec:ninn}. Multiple scattering will be reduced by employing thinner sensors as well as thinner read-out chips. At the moment within the ATLAS Planar Pixel Sensor R\&D Project sensors with thicknesses between (75--150)\,\mum{} are being investigated. Of key interest here is the ratio of collected charge to the threshold of the read-out chip. The new IBL read-out chip FE-I4 offers good operational performance down to thresholds of about 1\,ke \cite{MaltePsD}, while the FE-I3 can only be used down to around 3\,ke. For the read-out chip to be developed for the inner layers a low threshold will again be a key requirement, such that it is compatible with the collected charges discussed in the following. 

\subsubsection{Slim Edges}
\label{sec:slimedge}
Slim edges, i.\,e.\ a reduced distance between the last pixel implant and the sensor edge, can be achieved in different ways. Optimising the guard ring layout and reducing the safety margin  already allowed to decrease the inactive edge from approximately 1.1\,mm, as in the currently used  ATLAS sensors, down to 400\,\mum{} \cite{NinPpaper,Tobidice}. However, in the context of the phase II upgrade further reduction is due. In the following, four approaches are discussed: three rely on a treatment of the edge, while the last is design based.

The first approach employs Deep Reactive Ion Etching (DRIE) \cite{Boschpatent} to achieve trenches around the sensors, which allow for a doping of the sensor sides. The two dedicated productions using this approach rely on n-in-p sensors. In one production designs by the MPP/HLL and LAL groups are implemented on wafers of 100\,\mum{} and 200\,\mum{} thickness employing the VTT production lines \cite{AnnaJapan,JuhaJapan}. To achieve the interpixel isolation the p-spray isolation method was transferred from HLL to VTT.  Here the sides are implanted slanted with boron to define the electrical field on the sensor side. The edge distance is reduced by implementing only one guard and/or bias ring instead of a full guard ring scheme. By this, the edge distance was reduced down to 50\,\mum{}. The modules were interconnected to FE-I3 and FE-I4 read-out chips using the bump-bonding technique by VTT. Here, the UBM and solder bumps were applied by IZM for the FE-I3 read-out chips, and by VTT for the FE-I4 read-out chips. All sensors underwent the UBM preparation at VTT. In laboratory measurements with radioactive sources, within uncertainties, the same MPV of the collected charge was observed for the central and the edge pixels \cite{AnnaJapan}. This indicates that the edge region is actively contributing to the charge collection. For a determination of the hit efficiency beam test measurements are being conducted at the moment.  

\begin{figure}[!b]
\centering
\includegraphics[width=0.98\columnwidth]{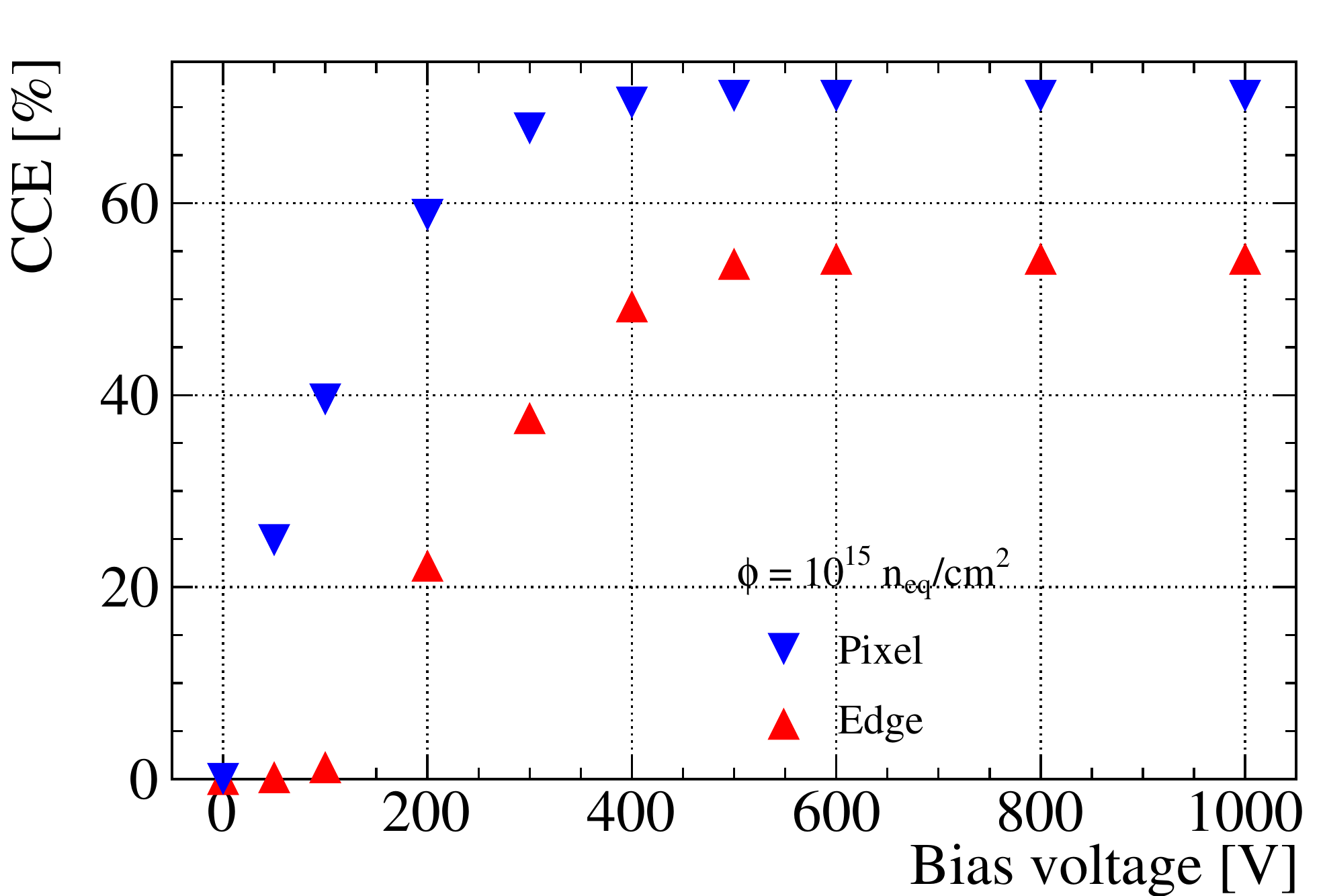}
\caption{Simulated CCE for a laser pulse as a function of the applied bias voltage for a slim edge sensor produced using the LPNHE/FBK approach. The blue triangles indicate the CCE for a injection below a pixel implantation, the red triangles for a injections outside of the guard ring structures. The fluence simulated is $10^{15}$\,\neqcm.}
\label{fig:FBK}
\end{figure}

In another production employing DRIE etching, FE-I3 and FE-I4 read-out chip compatible designs by the groups from LPNHE and FBK are being produced on the FBK production line using wafers of 200\,\mum{} thickness \cite{LPNHEFBK1}. The trench is doped by diffusion, as it is used for the production of 3D sensors \cite{Parker3D}. The typical edge distance achieved is between 100\,\mum{} and 200\,\mum{}, by reducing the guard ring scheme considerably. 
To predict the behaviour after irradiation, infra-red laser injections into a sensor irradiated to a fluence of $10^{15}$\,\neqcm{} were simulated. One injection was simulated to occur outside of the guard ring scheme, i.\,e.\ in the edge region, the other was simulated in the area of a pixel implant.  In Figure\,\ref{fig:FBK} the charge collection efficiency (CCE) with respect to the simulated pre-irradiation value is shown as a function of the applied bias voltage for both simulation points.
As expected, higher bias voltages are needed to deplete the region under the guard rings than for the one under the pixel implantations. Above 500\,V a saturation of the CCE is found around 55\,\% (70\,\%) in the edge (pixel) region. The predicted CCE at the edge is slightly lower than below the pixel implant since the traversed path is longer and thus trapping becomes more important. Assuming a low threshold operation, that is possible with the FE-I4 read-out chip, the results indicate a good performance of the devices. 

Another approach to achieve slim edges is the Scribe-Cleave-Passivate Approach \cite{SCP}, developed by the SCIPP group in collaboration with the U.S. Naval Research Laboratory (NRL). As a post processing step, it allows to achieve slim edges also for already produced sensors with a traditional design. It relies on a low damaged side wall, which is achieved by cleaving along a scribe line, defined by a simple lithographic step. The exposed side-walls are then passivated using atomic layer deposition of alumina for n-in-p sensors and of SiO$_2$ or Si$_3$N$_4$ for sensors with an n-type bulk. This leads to a controlled drop of the potential along the side-wall. The approach was successfully tested with sensors from several different productions. Further details can be found in \cite{SCP,TobiJapan}. 

\begin{figure}[ht]
\centering
\includegraphics[angle=90,width=\columnwidth]{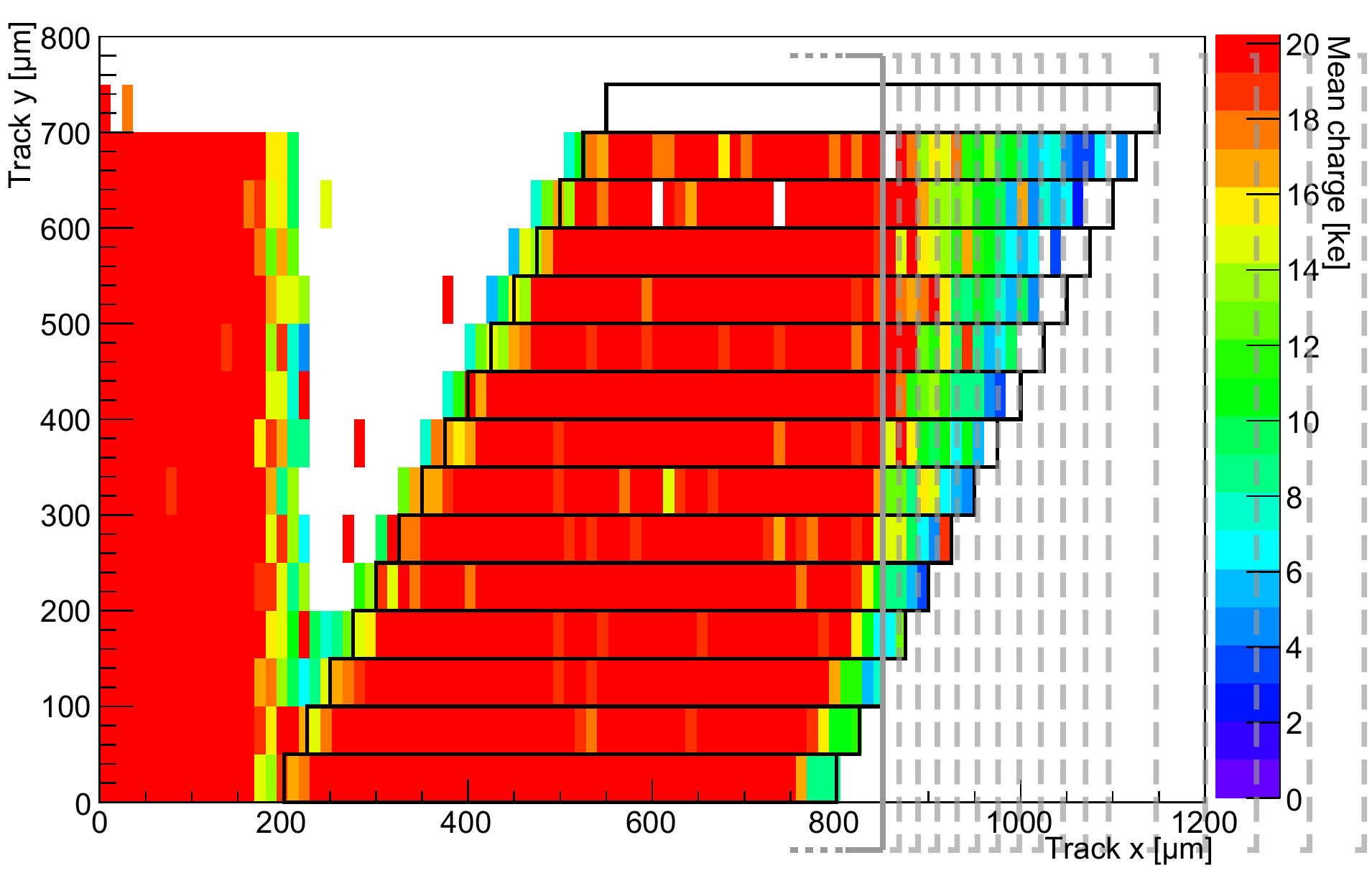}
\caption{Mean collected charge as a function of the impact point predicted by the beam telescope for an FE-I3 module with pixel implantations, which are shifted step-wise beneath the guard ring structure on the opposing sensor side \cite{PPSTB}. Each group is projected into a single pixel.}
\label{fig:DoStep}
\end{figure}

In the last approach to achieve slim edges the guard rings are shifted beneath the pixel implantation, and thus this is only possible for sensors, were the guard rings are placed on the opposite side of the pixel implants, namely in n-in-n sensors. The minimum edge distances achievable are around 200\,\mum. To investigate this approach, a special design was included in a production at CiS by the TU Dortmund group \cite{TobiJapan,BenoitPhD,WittigPhD}, which exhibits step-wise shifted pixel implantations, i.\,e.\ the pixel implantations at the sensor edges were shifted below the guard rings in groups using different distances. The such built modules were tested in a beam test at the CERN SPS with 120\,GeV pions. In Figure\,\ref{fig:DoStep} the pixels for each group are overlaid, and the mean collected charge as a function of the impact position predicted by the beam telescope is shown. The position of the guard rings are indicated in grey. 
Although the electric field is less homogeneous in this region the charge collection is increased significantly, especially when compared to the threshold and the. Following this results, the approach was adopted for the planar sensors employed in the IBL \cite{IBL_Proto}.

\subsubsection{Performance of n-in-n Pixel Detectors}
\label{sec:ninn}
As a candidate for the inner layers the performance of n-in-n sensors is investigated up to $2\cdot10^{16}$\,\neqcm{} using sensors from productions at CiS employing designs by the TU Dortmund group \cite{WittigPhD}. The sensors have thicknesses of 250\,\mum{} or 285\,\mum{} and were interconnected to the ATLAS FE-I3 read-out chip with two different technologies. In the first approach solder bump bonding by IZM was used, in the second approach low temperature indium bump bonding was applied, which allows for an interconnection after irradiation of the sensor, while minimizing annealing effects. In Figure\,\ref{fig:CisNinPnirrad} the MPVs of the collected charge for the assembly irradiated to $5\cdot10^{15}$\,\neqcm{} are given. At 1\,kV the MPV of the collected charge is 10.3\,ke as determined for $\beta$-electrons originating from a \Sr-source, i.\,e.¸\ well above the threshold of 3.2\,ke. After an irradiation to the highest fluence of $2\cdot10^{16}$\,\neqcm{} an MPV of the collected charge of 4.2\,ke is measured at an applied bias voltage of 1\,kV. Although this is low when compared to the typical thresholds of the FE-I3 read-out chip, it is well above threshold when employing an FE-I4 read-out chip as discussed above.

In beam test measurements at the CERN SPS with 120\,GeV pions and at DESY with 4\,GeV positrons, the hit efficiency was determined as a function of the applied bias voltage and the received fluence. 
At a bias voltage of 1\,kV a preliminary hit efficiency of $95.4$\,\% at a fluence of $10^{16}$\,\neqcm{} and of 88.4\,\% at a fluence of $2\cdot10^{16}$\,\neqcm{} was determined. For all fluences an increase with applied bias voltage is found, and a high hit efficiency can be regained, when applying a higher bias voltage up to 1.8\,kV. When employing the FE-I4 read-out chip the bias voltage requirements become less stringent, given the lower possible threshold \cite{IBL_Proto,TobiJapan}. Anyhow, in laboratory environment the modules can be operated stably at these high bias voltages. As for the HPK and MPI-HLL modules the main losses occur in the region of punch through biasing and in the corners. So all quoted hit efficiencies  have to be taken as lower bound, since inclined tracks are less affected by this. Further results are summarised in \cite{RummlerPhD,LapsienDiploma}.

\subsubsection{Design Improvements}
To overcome the efficiency losses in the punch-through structure for perpendicular impinging particles, different design modifications are under investigation. The KEK group in collaboration with HPK replaced the punch through biasing with a poly-silicon resistor, encircling the pixel implant \cite{UnnoHiro}. In another production by the TU Dortmund group in collaboration with CiS the routing of the metallization of the bias grid is altered, aiming at a different electric field configuration \cite{WittigPhD}. First results are expected shortly.

\section{Conclusion \& Outlook}
Planar pixel sensors are a well understood and established technology, which exhibits excellent performance within the present tracking detectors of the ATLAS and CMS experiments. The latest results obtained by the ATLAS Planar Pixel Sensor R\&D Project and presented here imply a good performance also after the high irradiation levels  expected after the upgrades of the LHC accelerator complex. Furthermore, they offer the cost-effectiveness needed for the large instrumented areas foreseen in the upgrades of the pixel systems of the LHC experiments. 

Especially, it has been shown that the hit efficiency after HL-LHC inner layer fluences is sufficient, if high enough bias voltages are applied. Using n-in-p sensors of the same thickness as the currently used detectors, already at a moderate bias voltage of 600\,V efficiencies above 97\,\% were observed after a received fluence of $10^{16}$\,\neqcm. For n-in-n sensors fluences up to $2\cdot10^{16}$\,\neqcm{} were explored and high hit efficiencies of 97.5\,\% were found at a bias voltage of 1.5\,kV.

Different productions incorporating slim edges and/or implanted side walls, were discussed. For the three already finished ones a good performance was found, even with a minimal edge distance of 50\,\mum{}.

\section{Acknowledgements}
\label{sec:acknowledgment}
This work has been partially performed in the framework of the CERN RD50 Collaboration. The authors thank A.~Dierlamm (KIT), V.~Cindro, and I.~Mandić (Jo\v{z}ef-Stefan-Institut) for the sensor irradiations. Part of the irradiation were supported by the Initiative and Networking Fund of the Helmholtz Association, contract HA-101 ("Physics at the Terascale"). Another part of the irradiations  and the beam test measurements leading to these results has received funding from the European Commission under the FP7 Research Infrastructures project AIDA, grant agreement no. 262025.

\end{document}